\documentclass[aps,prl,twocolumn,showpacs,superscriptaddress,groupedaddress]{revtex4-1}

\usepackage{multirow}
\usepackage{mathrsfs,amsmath}
\usepackage{hyperref}
\usepackage{verbatim}
\usepackage{amsmath}
\usepackage{amsfonts}
\usepackage{amssymb}
\usepackage{amsthm}
\usepackage{bm}
\usepackage{xparse}
\usepackage{graphicx}
\usepackage{xcolor}
\usepackage{textcase}
\usepackage{url}
\usepackage{lipsum}
\usepackage{appendix}
\usepackage{epstopdf}
\usepackage{footnote}
\usepackage{color}

\newcommand{\ket}[1]{| {#1} \rangle} % for Dirac bras
\newcommand{\bra}[1]{\langle {#1} |} % for Dirac kets
\DeclareDocumentCommand{\Tr}{m m O{\big}}{{\rm Tr}_{\:\!{#1}}#3({#2}#3)}

\begin{document}
\title{Two-way communication with a single quantum particle}
\author{Flavio Del Santo}
\affiliation{
Faculty of Physics,
University of Vienna,
Boltzmanngasse 5,
A-1090 Vienna, Austria}
\author{Borivoje Daki\'c}
\affiliation{
Institute for Quantum Optics and Quantum Information (IQOQI),
Austrian Academy of Sciences, Boltzmanngasse 3,
A-1090 Vienna, Austria
}
\affiliation{
Vienna Center for Quantum Science and Technology (VCQ), Faculty of Physics, Boltzmanngasse 5, University of Vienna, Vienna A-1090, Austria
}

\date{\today}

\begin{abstract}
In this letter we show that communication when restricted to a single information carrier (i.e. single particle) and finite speed of propagation is fundamentally limited for classical systems. On the other hand, quantum systems can surpass this limitation. We show that communication bounded to the exchange of a single quantum particle (in superposition of different spatial locations) can result in ``two-way signaling'' which is impossible in classical physics.
%We derive a ``no-go'' theorem and quantify the discrepancy between classical and quantum scenario by the win probability of game played by distant players.
%Our result brings novel insight and opens new possibilities in the field of quantum communication.
We quantify the discrepancy between classical and quantum scenarios by the probability of winning a game played by distant players. We generalize our result to an arbitrary number of parties and we show that the probability of success is asymptotically decreasing to zero as the number of parties grows, for all classical strategies. In contrast, quantum strategy allows players to win the game with certainty.
\end{abstract}

\maketitle

\subsection{Introduction}
Generally speaking, communication is the process of transmitting a message (information) from a sender to a receiver~\cite{Shannon}. We usually think of sending physical information, i.e. a message embodied in an information carrier and sent as a signal, such as voltage signals, speech, video or radar. In the classical world, physical systems that carry information obey the laws of classical physics. For example, electromagnetic signals propagate in space according to Maxwell's equations, thus the speed of information transfer is fundamentally limited to that of light. Similarly, in radio communications, information flows from a radio emitter to the radio receiver but not vice-versa, as it follows from the causality principle. In other words, abilities and limitations of communication and information processing in general are governed by the laws of physics. From that perspective, quantum physics together with its counterintuitive principles allows for novel possibilities that are not permissible in classical world. Encoding, transmitting and decoding information carried by quantum systems enables distant parties to beat the limits fundamentally imposed by the laws of classical physics. The most prominent examples include quantum communication complexity \cite{brassard2001quantum, buhrman2010nonlocality}, quantum key distribution \cite{bennett2014quantum, ekert1991quantum}, quantum teleportation \cite{bennett1993teleporting} and many others.

%In this respect, there is a plethora of examples, ranging from the celebrated CHSH communication game \cite{clauser1969proposed} to the very recent proposal of ``quantum acausal processes'' \cite{oreshkov2012quantum}, superposition of orders \cite{feix2015quantum} and directions \cite{guerin2016exponential}, quantum combs [Ref], quantum switch \cite{chiribella2013quantum}, quantum causal models \cite{allen2016quantum} that are in some sense related to our work. Some of these novel phenomena have been demonstrated in recent experiments \cite{procopio2015experimental, rubino2017experimental}.

In this letter we further investigate the discrepancy between classical and quantum information processing. We analyze the model of communication that is restricted to a) \emph{a single information carrier} (i.e. single particle), and b) \emph{the finite speed of propagation}. We show that the model suffers from fundamental limitations when restricted to classical systems (particles). On the other hand, quantum mechanics allows for a novel possibility, i.e. to put particles in superposition of spatially distinct locations, which turns out to be a more powerful resource for communication (as compared to its classical counterpart). In particular, we show that communication restricted to the exchange of a single quantum particle that is coherently distributed at different spatial locations can result in a two-way signalling, which is essentially impossible by using classical resources. Based on the model, we introduce a simple game played by distant players that are supposed to accomplish a certain task by exchanging (limited) communication. We show that the probability to win the game is  bounded and strictly lower than 1 for all classical strategies. In contrast, quantum information carriers in (spatial) superpositions enable players to accomplish the task with certainty. Unlike many quantum information protocols based on entanglement and quantum correlations between different parties, our task involves only a single quantum particle and it is based solely on the superposition principle. In this respect our findings are similar to recent proposals exploring quantum superpositions for information processing purposes, such as ``quantum acausal processes'' \cite{oreshkov2012quantum}, superposition of orders \cite{feix2015quantum} and directions \cite{guerin2016exponential}, quantum combs~\cite{Chiribella}, quantum switch \cite{chiribella2013quantum} and quantum causal models \cite{allen2016quantum}. Some of these novel phenomena have been demonstrated in recent experiments \cite{procopio2015experimental, rubino2017experimental}.

%%%%%%%%%%%%%%%%%%%%%%%%%%
\begin{figure}[]
\centering
\includegraphics[width=8cm]{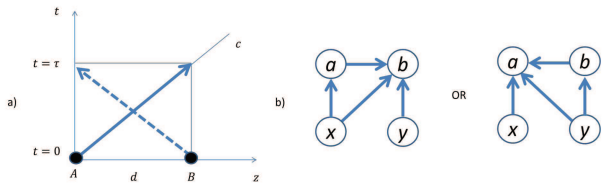}
\caption{{\bf a) Space-time diagram.} \small{Within the time window $\tau$ either $A$ signals to $B$ or $B$ to $A$.} {\bf b) Causal diagram.} \small{There are two possible causal relations between variables $x$, $y$, $a$, $b$, i.e. either $x$ influences $a$ and $b$, $a$ influences $b$ (in general, $a$ can be generated in past of $b$, i.e. at time $t=0$), whereas $y$ influences $b$ only (left) or vice-versa (right).}}
\label{fig:1}
\end{figure}
%%%%%%%%%%%%%%%%%%%%%%%%%%
\subsection{Two-way communication with a single particle}
Consider a classical model of communication where two agents Alice and Bob are located at a distance $d$ from each other and they are allowed to communicate via a single information carrier. Here as a carrier, we think of a particle or object that can travel with the finite speed bounded by $c$. For simplicity, we assume that the speed of information transfer matches the maximal value $c$. For example, if Alice holds the particle, she can imprint the message in it and send it to Bob. The message needs $d/c$ time to arrive at Bob's side. We assume that the communication channel is open for a certain time window of $d/c\leq\tau\leq d/c+\epsilon$, where $\epsilon\geq0$ is a small constant. The time window $\tau$ is set such that the particle has enough time to arrive at Bob's side, but not to come back to Alice. This is what we mean when we refer to limited communication, i.e. within the time window $\tau$, $A$ and $B$ can exchange only a ``one-way'' communication. At time $t=0$, $A$ and $B$ are given the input variables $x$ and $y$ by the referee and they are asked to reveal the output variables $a$ and $b$ at the later time $t=\tau$. If we represent the communication in the space-time diagram (see Figure 1a), it is clear that there are only two possible options, i.e. if the particle was in possession of Alice at $t=0$, she can encode her input in it and send it to Bob, but she gets no information on Bob's input, or vice-versa. In the formalism of causal diagrams \cite{allen2016quantum}, there are two possible causal relations between the variables $x$, $y$, $a$, $b$ as shown in Figure 1b. Therefore, the probability distribution $p(ab|xy)$ is a classical mixture of one-way signaling distributions, i.e.
\begin{eqnarray}\label{class dist}\nonumber
p(ab|xy)&=&\lambda p_{A}(a|x)p_{A\prec B}(b|xya)\\
&+&(1-\lambda)p_{B}(b|y)p_{B\prec A}(a|xyb),
\end{eqnarray}
where symbol $\prec$ denotes the direction of signaling, e.g. $A\prec B$ denotes the case of $A$ sending her particle to $B$. The probability distribution \eqref{class dist} is completely characterized by a the ``so-called'' classical polytope \cite{brunner2014bell}, and its facets are represented by the Bell's-like inequalities which impose the limits on classical models. For the case of binary inputs $x,y=0,1$ and binary outputs $a,b=0,1$, there are only two inequivalent inequalities \cite{branciard2015simplest}
\begin{eqnarray}\label{GYNI}
p_{}(a=y,b=x)&\leq&\frac{1}{2},\\\label{LGYNI}
p_{}(x(a\oplus y)=y(b\oplus x)=0)&\leq&\frac{3}{4},
\end{eqnarray}
known as two variants of ``guess  your  neighbor's  input'' (GYNI) game~\cite{almeida2010guess}. Here $x$ and $y$ are uniformly distributed, i.e. $p(x,y)=1/4$. We will focus on \eqref{GYNI} which, when translated into the language of communication games, results in the requirement of computing the neighbor's  input. More precisely, for given inputs $x$ and $y$, $A$ and $B$ are asked to reveal the outputs $a=y$ and $b=x$, respectively. Clearly, in the classical scenario, the probability of success is bounded by \eqref{GYNI}.

%Before we proceed further, we will re-formulate the rules of the game introduced above. Namely, for the sake of simplicity, instead of GYNI game, the players are asked to guess the parity of the inputs, i.e. to output $a=b=x\oplus y=p$. Note that this is completely equivalent to the GYNI condition ($a=y$ and $b=x$) as if the players know the parity, they can easily extract the value of the neighbor's input, i.e. $y=p\oplus x$ for $A$, and $x=p\oplus y$ for $B$ and vice-versa.

%Assume also that the players, fully aware of the rules, are granted an \textit{initialization phase} in which they can exchange unlimited information and agree on a strategy. The referee challenges the players to guess the parity of the initial inputs, i.e.  $ f(x,y)= x  \oplus y$,  where $ \oplus$ denotes the sum modulo-2. Therefore, the players win the game if and only if the \emph{predicate} $a=b= x  \oplus y$ is fulfilled. In a classical scenario, they previously agree that one of them, say Alice, will always send the value of her input to Bob. In such a way, Bob knows both the initial inputs and is always able to guess the parity. On the contrary, Alice has only a partial information and she can merely guess the right result of $f(x,y)$ with probability 1/2. Thus, in the event that the referee asks, with uniformly distributed probability, only one of the two players to answer, then the win probability is bounded by $P_{class}'\leq3/4$. Otherwise, if both the players are asked to guess the parity, they win the game with probability $P_{class}\leq1/2$.

%%%%%%%%%%%%%%%%%%%%%%%%%%
\begin{figure}[]
\centering
\includegraphics[width=8cm]{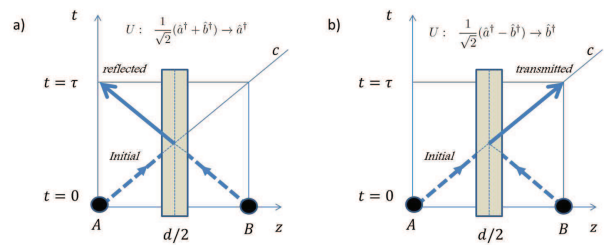}
\caption{{\bf Communication with a single quantum particle.} The particle is prepared in superposition of two spatial locations $A$ and $B$. At time $t=0$ Alice and Bob encode their inputs and send their ``parts of the particle'' to the partner. On the way, the particle hits the unitary device (e.g. potential barrier) and gets ``half-reflected'' and ``half-transmitted'' in a coherent way (see main text). {\bf a)} If the parity of inputs is $s=0$, the transformation $U$ deterministically guides the particle to the Alice's location. {\bf b)} An analogous situation where $s=1$ and the particle deterministically ends at Bob's location.}
\label{fig:2}
\end{figure}
%%%%%%%%%%%%%%%%%%%%%%%%%%

In contrast to the classical case, quantum mechanics allows to put the particle in superposition of different spatial locations. Let $A$ and $B$ share a single quantum particle prepared in a superposition of their two distinct locations, i.e. $\propto\ket{\mathrm{particle~with}~A}+\ket{\mathrm{particle~with}~B}$.  For the sake of simplicity of notation, we will use the second quantization formalism. Thus, the initial state of the system is given by
\begin{equation}
\label{initial}
|\psi\rangle_{in} = \frac{1}{\sqrt{2}}  (|0\rangle_A |1\rangle_B + |1\rangle_A |0\rangle_B) = \frac{1}{\sqrt{2}} (\hat a^\dagger + \hat b^\dagger)|0\rangle_A |0\rangle_B,
\end{equation}
where, for example $|1\rangle_A|0\rangle_B$ indicates that particle is localized with Alice, whereas Bob has zero (vacuum) particles in possession. The operators $\hat a^\dagger$ and $\hat b^\dagger$ are the standard ladder operators that create the particle on $A$'s and $B$'s side, respectively. Note that we use ladder operators just for convenience, and as long as we are dealing with a single particle, our results are completely independent of the distinguishability property or the type of particle used. After receiving their inputs at time $t=0$, the players will encode them by adding a local phase to the state, i.e. $\hat a^\dagger \xrightarrow{}  (-1)^{x}\hat  a^\dagger$ and $ \hat b^\dagger \xrightarrow{}  (-1)^{y} \hat b^\dagger$. Thus, the initial state is transformed into
\begin{eqnarray}\label{encoded}
\ket{\psi}_{in}&\xrightarrow{}&\frac{1}{\sqrt{2}} ((-1)^{x}\hat a^\dagger + (-1)^{y}\hat b^\dagger)|0\rangle_A |0\rangle_B
%&=&\frac{1}{\sqrt{2}} (-1)^{x}(\hat a^\dagger + (-1)^{p}\hat b^\dagger)|0\rangle_A |0\rangle_B
\end{eqnarray}
Immediately after the encoding (practically at time $t\approx 0$), each player sends its ``part of the particle'' to the partner. In addition, a unitary device is placed right in the middle between $A$ and $B$ (at the distance $d/2$, see Figure \ref{fig:2}), such that the particle, when sent from $A$ to $B$, is ``half-reflected'' and ``half-transmitted'' in a coherent way. A similar situation is found if the particle is sent from $B$ to $A$. The unitary device serves as a communication channel, and it can be realized in practice by putting a simple potential barrier for material particles or an ordinary beam-splitter for the case of single-photons. In other words, the unitary device transforms the modes in the following way
%\begin{eqnarray}\label{comm channel}
%\hat a^\dagger &\xrightarrow{}& \frac{1}{\sqrt2}(\hat  a^\dagger +\hat  b^\dagger),\\
%\hat b^\dagger &\xrightarrow{}& \frac{1}{\sqrt2}(\hat  a^\dagger -\hat  b^\dagger).
%\end{eqnarray}
\begin{equation}\label{comm channel}
\hat a^\dagger \xrightarrow{} \frac{1}{\sqrt2}(\hat  a^\dagger +\hat  b^\dagger),~~\hat b^\dagger \xrightarrow{} \frac{1}{\sqrt2}(\hat  a^\dagger -\hat  b^\dagger).
\end{equation}
Finally, after exchanging the communication (at time $\tau$), the final state reads
\begin{eqnarray}
\ket{\psi}_f
=\left\{
   \begin{array}{ll}
     \ket{1}_A\ket{0}_B, & \hbox{$x=0,~y=0$;} \\
     \ket{0}_A\ket{1}_B, & \hbox{$x=0,~y=1$;} \\
     -\ket{0}_A\ket{1}_B, & \hbox{$x=1,~y=0$;} \\
     -\ket{1}_A\ket{0}_B, & \hbox{$x=1,~y=1$.}
   \end{array}
 \right.
\end{eqnarray}
At the last stage, the players perform the measurement, and $A$ will find the particle in her possession whenever the parity of the inputs $s=x\oplus y=0$, whereas the particle is located at Bob's side for $s=1$. Therefore, the players can read the parity of inputs perfectly, from which they can easily extract the value of the neighbor's input, i.e. $a=s\oplus x=y$ for $A$, and $b=s\oplus y=x$ for $B$. Thus they win the game with certainty. Since $A$ and $B$ beat the bound \eqref{GYNI} it is clear that the resulting probability distribution $p(ab|xy)$ is a two-way signaling distribution.

\subsection{$N$-partite game}
Let us consider $N$ players $A_0,\dots,A_{N-1}$ located at the vertices of a regular polygon each of them at a distance $d/2$ from the polygon center. Here we set $N$ to be a prime number. As before, the players share a single particle that can be communicated during the time window $d/c\leq\tau\leq d/c+\epsilon$. Furthermore, let us assume that the particle has to be communicated through the center of polygon (we will analyze a more general situation by the end of this chapter) so that it has enough time to be exchanged between two players at most. At $t=0$ the players are given the set of inputs $x_0,\dots,x_{N-1}\in\{0,\dots,N-1\}$ and they are asked to reveal the binary outputs $a_0,\dots,a_{N-1}\in\{\mathrm{YES},\mathrm{NO}\}$ at time $t=\tau$. The input string $X=(x_0,\dots,x_{N-1})$ is defined by two integers $(n,m)$, i.e. $x_k=nk+m ~(\mathrm{mod}~N)$, with $(n,m)$ picked randomly from the set $n,m=0,\dots,N-1$ (with probability $1/N^2$). The referee asks each player, say $A_k$, to answer the following question:
\begin{itemize}
\item[$Q_k:$] \emph{Does the given input string $X$ satisfies $n=k$?}
\end{itemize}
After exchanging communication, at time $\tau$ the players are supposed to reveal their answers.
%, i.e. the $k$th player assigns $a_k=0/1$ for ``YES/NO'' answer, respectively.
In order to win the game, only one player has to answer with ``YES'' ($n=k$ is true for one $k$ only), while the rest shall answer with ``NO''. Our goal here is to show that the best classical strategy reveals at most $1/N$ chance to win the game, whereas quantum strategy allows players to win the game with certainty.

Before we proceed further, let us examine the simplest example of $N=2$. In such a case, the set of inputs reduces to $x_0=m$ and $x_1=n+m$, where $n,m=0,1$ are randomly picked with probability $1/2$. Clearly, the parity of inputs satisfies $s=x_0+x_1=n ~(\mathrm{mod}~2)$. The set of questions reduces  to ``$Q_0:$ Is $s=0$?'' and ``$Q_1:$ Is $s=1$?'', which is completely equivalent to the bipartite GYNI game discussed in the previous chapter, with the classical probability of success of $1/2$. Interestingly, even without communication the players achieve the same probability of success (e.g. they agree to output $s=0$ always). The reason for that are the rules of the game, i.e. they win the game iff they both posses the information on parity $s$ which is the joint property of inputs. Therefore, one round (one-way) communication does not help to increase the probability of success. However, as we will see later, for the case of $N$-partite game, more rounds of communication, indeed will help to boost the probability of success.

In the classical case, it is clear that within the time window $\tau$ only a single ``one-way'' communication between two players can occur (during $\tau$ the particle can cover distance $d$, thus it can travel from one vertex to the polygon center and from there to an arbitrary vertex). For example $A_k$ can keep the particle at $t=0$ and send his input to $A_l$. Now, since $A_l$ has in possession $x_k=n k+m$ and $x_l=n l+m$, he can simply find the difference $x_k-x_l=n(k-l)$ and from there he can extract the value of $n=(x_k-x_l)(k-l)^{-1}$ ($N$ is chosen to be a prime number, thus the division modulo $N$ is well defined). Therefore, $A_l$ can verify $n=l$ with certainty. Nevertheless, the rest of $N-1$ players do not have any information on inputs of the other players. Therefore, they have to check $n=k$ solely on their inputs $x_k=nk+m$ (here $k\neq l$). Since they are missing the information on $m$, they can only guess the value of $n$ (in order to validate $n=k$). One of the best strategies for them is simply to answer ``NO'' always. In this way they can achieve $1/N$ probability to win the game. The other potential strategy is to fix one player who will always reveal ``YES'' and the other ones shall answer with ``NO''. Again they achieve the same chance of $1/N$ to win the game. Interestingly, in this particular case, they do not have to exchange any communication.
%The best strategy is to agree on a particular value of $m$ in advance (e.g. they agree to set the value of $m$ before the game starts), and calculate $n=(x_k-m)k^{-1}$. Since the unknown variable $m$ is picked by referee randomly from the set $\{0,\dots N-1\}$, the best achievable probability to win the game is $1/N$.

Now we turn to quantum strategy. A single particle is prepared in equally weighted superposition of $N$ different locations, i.e. $\ket{\psi}_{in}=\frac{1}{\sqrt N}\sum_{k=0}^{N-1}\hat a_k^{\dagger}\ket{0}_0\dots\ket{0}_{N-1}$. Here, $\hat a_k^{\dagger}$ labels the operator creating the particle at $k$th player's location, i.e. $\hat a_k^{\dagger}\ket{0}_k=\ket{1}_k$. After getting their inputs, the players encode them into local phases $\hat a_k^{\dagger}\xrightarrow{}\omega^{x_k}\hat a_k^{\dagger}$, where $\omega=e^{2\pi i/N}$ and $x_k=n k+m$. In the next step, each of them sends his/her ``part'' of particle towards the polygon center, where the particle hits the unitary device that coherently ``splits'' the particle into $N$ different direction
\begin{equation}\label{U transf}
\hat a_k^{\dagger}\xrightarrow{}\sum_{l=0}^{N-1}U_{kl}\hat a_l^{\dagger}.
\end{equation}
Here $U$ is a discrete Fourier transformation, i.e. $U_{kl}=\frac{1}{\sqrt N}\omega^{-kl}$. In other words, the particle sent from $A_k$ to the polygon center is partially reflected (with probability $1/N$) and goes back to the sender, and it is partially scattered into $N-1$ different directions (each with probability $1/N$) pointing towards other $N-1$ players (vertices of the polygon). The simple calculation shows that the final state (after transformation) is given by
\begin{equation}
\ket{\psi}_f=\omega^m \hat a_n^{\dagger}\ket{0}_0\dots\ket{0}_{N-1}=\omega^m\ket{0}_0\dots\ket{1}_n\dots\ket{0}_{N-1}.
\end{equation}
Thus, at time $t=\tau$, only the $n$th player will find the particle in his possession. He is the only one who shall answer with ``YES'' to the posed question $Q_n$. The rest of players do not find the particle located at their positions, therefore they shall answer with ``NO''. In this way they win the game with certainty.

A slightly more general situation one finds if the communication is not bounded to go through the center of polygon, e.g. players can send the particle in any possible direction. Since the particle travels with the speed of $c$, during the time window $\tau=d/c$ it can pass the distance of at most $d_{m}=d$. Assume that the particle was located at $A_k$ at $t=0$. In such case, the optimal classical strategy of exchanging as much information as possible between players is to send the particle along the edges of a polygon. In this way, up to $k_{\mathrm{max}}$ players will acquire information on the input string $X$, where $k_{\mathrm{max}}=\lfloor 1/ \sin{\frac{\pi}{N}}\rfloor$. Here $\lfloor . \rfloor$ denotes the integer part function. Nevertheless, the remaining $N-k_{\mathrm{max}}$ players will have no information about $X$. As before, they can choose to output ``NO'' always. In such a case, they achieve the probability of success $k_{\mathrm{max}}/N= \lfloor1/\sin{\frac{\pi}{N}}\rfloor/N   \xrightarrow{}\frac{1}{\pi}$, when $N\xrightarrow{}+\infty$. Recall that in a single round of one-way communication (particle traveling through the polygon center), only one player gets the full knowledge on the input string $X$, thus the winning probability is $1/N$. As we have shown, this bound is achievable even without communication. However, in this case where more players get the full knowledge on $X$, the probability of success grows with the number of communication rounds $k_{\max}\geq2$ and the players start to benefit from communication. A simple calculation shows that the minimal number of players needed to exceed the threshold is $N\geq7$ ($N$ is prime).

\subsection{Fock space perspective}

The Fock space formulation presented here gives a good way to analyze the protocol from the information-theoretic perspective. We examine $N=2$ in details; the generalization to $N>2$ is straightforward. Firstly, we divide the protocol into three parts: a) encoding of inputs into a resource state, b) transmission through the communication channel, and c) decoding. Classically, the inputs are encoded into memory of an information carrier (particle), which we label via the set of states $\Sigma=\{\bot,a_1,a_2,...\}$. Here $\bot$ denotes the ``vacuum state'', which formally captures the situation of  no-particle present. For arbitrary encoding function $E$, we have $E(\bot)=\bot$, meaning that no-information can be stored in vacuum. Classically, a resource state $\rho_{ab}$ is a probability distribution (shared randomness) on $\Sigma\times \Sigma$. Since Alice and Bob share a single carrier, there is a restriction on $\rho_{ab}$, i.e. we have a non-zero support for $\rho_{\bot b}$ and $\rho_{a\bot}$ only (meaning, the carrier is located at $A$ or $B$). Therefore, in the most general case, $\rho_{ab}$ is as a mixture $\rho_{ab}=\lambda\delta_{\bot,a}r_b+(1-\lambda)\delta_{\bot,b}r_a$, where $r_a,r_b$ are some distributions on $\Sigma$. Now, for given inputs $x$ and $y$, the players choose certain encodings $A_x,B_y:\Sigma\mapsto\Sigma$, which map the resource state into $\tilde{\rho}_{ab}=\lambda\delta_{\bot,a}r_{B_y(b)}+(1-\lambda)\delta_{\bot,b}r_{A_x(a)}=\lambda \rho^{(y)}+(1-\lambda) \rho^{(x)}$. Thus, in every single run, the encoded state contains information on one input only (either $\rho^{(x)}$ or $\rho^{(y)}$), but never both of them simultaneously, for arbitrary encoding. Here we used $A_x(\bot)=\bot$ and $B_y(\bot)=\bot$. Finally, after passing $\tilde{\rho}$ trough a (memoryless) communication channel $C$, we get the final state $\rho_{out}=\lambda C(\rho^{(y)})+(1-\lambda)C(\rho^{(x)})$, which means that the final distribution (after decoding) is a mixture of ``one-way'' signalling distributions given by the equation \eqref{class dist}.

On the other hand, the resource state in quantum scenario is a ``coherent mixture'' given by \eqref{initial}. After encoding, the global state \eqref{encoded} contains the information on both inputs simultaneously, which in turn enables ``two-way'' signaling (after sending the state through a suitable communication channel \eqref{comm channel}). Similarly to the classical case, zero information can be stored in vacuum state, i.e. $U\ket{0}=\ket{0}$, for any encoding unitary $U$. Nevertheless, the possibility of having a resource state in superposition of two classical alternatives (i.e. $\rho_{\bot b}$ and $\rho_{a\bot}$) essentially enables for quantum advantage.

\subsection{Resistance to noise}
Let us consider the input state $\rho=(1-\lambda)\ket{\psi}_{in}\bra{\psi}+\lambda \rho_{noise}$, where $\ket{\psi}_{in}=\frac{1}{\sqrt N}\sum_{k=0}^{N-1}a_k^{\dagger}\ket{0}_0\dots\ket{0}_{N-1}$ and $\rho_{noise}$ represents the noise, such as the white noise $\rho_{noise}^{(W)}=\frac{1}{N}\openone_N$ or the particle loss $\rho_{noise}^{(L)}=\ket{vac}\bra{vac}$~\cite{Vertesi12,Vertesi16}. Here $\ket{vac}=\ket{0}_0\dots\ket{0}_{N-1}$ is the global vacuum state and $\openone_N=\sum_{k=0}^{N-1}a_k^{\dagger}\ket{vac}\bra{vac}a_k$ is the identity in the single-particle subspace (this is a reasonable model for the white noise if there are no particle creations/anihilations). The overall probability of success reads $p_s=1-\lambda+\lambda p_{noise}$, where $p_{noise}\in[0,1)$ is the characteristic of noise. The classical bound for probability of success is $1/N$, therefore we get the upper noise threshold of $\lambda_c=\frac{1-\frac{1}{N}}{1-p_{noise}}$. If the white noise has been prepared, the transformation $U$ given by the equation $(8)$ (see main text) leaves $\rho_{noise}^{(W)}$ invariant, and we get $1/N$ chance for particle being detected at $A_k$; thus $p_{noise}=1/N$. We get $\lambda_c=1$, meaning that $\ket{\psi}_{in}$ is extremely tolerant to the white noise. Similarly, if $\rho_{noise}^{(L)}$ has been realized in the setup, none of the players will detect particle at his/her location ($\rho_{noise}^{(L)}$ is invariant under $U$ given by the equation $(8)$). In such a case, they all output ``NO'' and this is clearly a failure, hence $p_{noise}=0$. In this case, we get $\lambda_c=1-\frac{1}{N}$. In general, the upper noise threshold satisfies $\lambda_c=\frac{1-\frac{1}{N}}{1-p_{noise}}\geq1-\frac{1}{N}$ which is a remarkable resistance of $\ket{\psi}_{in}$ to arbitrary type of noise.

\subsection{Conclusions}
In this letter we have investigated the power of quantum superpositions for communication purposes. Interestingly, we have shown that a single quantum particle prepared in superposition of different spatial locations can be a stronger resource for communication (as compared to the classical counterpart). Furthermore, we show in the Appendix that our scheme shows a remarkable resistance to arbitrary type of noise. One may notice that our framework does not assume (\emph{a priori}) the use of quantum entanglement, in contrast to majority of known quantum information tasks and protocols. Nevertheless, when formulated in the second quantization language, our resource state $\ket{\psi}_{in}=\frac{1}{\sqrt N}\sum_{k=0}^{N-1}a_k^{\dagger}\ket{0}_0\dots\ket{0}_{N-1}$ is exactly the $W$-state~\cite{Wstate} which is know to be a highly entangled state (particle-vacuum entanglement). Thus our results sheds new light on entanglement in Fock space (mode entanglement)~\cite{Vedral} and its applications for quantum information purposes.

\begin{acknowledgments}
\emph{Acknowledgments.}---The authors thank \v{C}aslav Brukner and \"{A}min Baumeler for helpful comments.
\end{acknowledgments}
\bibliography{library}

\begin{thebibliography}{21}%
\makeatletter
\providecommand \@ifxundefined [1]{%
 \@ifx{#1\undefined}
}%
\providecommand \@ifnum [1]{%
 \ifnum #1\expandafter \@firstoftwo
 \else \expandafter \@secondoftwo
 \fi
}%
\providecommand \@ifx [1]{%
 \ifx #1\expandafter \@firstoftwo
 \else \expandafter \@secondoftwo
 \fi
}%
\providecommand \natexlab [1]{#1}%
\providecommand \enquote  [1]{``#1''}%
\providecommand \bibnamefont  [1]{#1}%
\providecommand \bibfnamefont [1]{#1}%
\providecommand \citenamefont [1]{#1}%
\providecommand \href@noop [0]{\@secondoftwo}%
\providecommand \href [0]{\begingroup \@sanitize@url \@href}%
\providecommand \@href[1]{\@@startlink{#1}\@@href}%
\providecommand \@@href[1]{\endgroup#1\@@endlink}%
\providecommand \@sanitize@url [0]{\catcode `\\12\catcode `\$12\catcode
  `\&12\catcode `\#12\catcode `\^12\catcode `\_12\catcode `\%12\relax}%
\providecommand \@@startlink[1]{}%
\providecommand \@@endlink[0]{}%
\providecommand \url  [0]{\begingroup\@sanitize@url \@url }%
\providecommand \@url [1]{\endgroup\@href {#1}{\urlprefix }}%
\providecommand \urlprefix  [0]{URL }%
\providecommand \Eprint [0]{\href }%
\providecommand \doibase [0]{http://dx.doi.org/}%
\providecommand \selectlanguage [0]{\@gobble}%
\providecommand \bibinfo  [0]{\@secondoftwo}%
\providecommand \bibfield  [0]{\@secondoftwo}%
\providecommand \translation [1]{[#1]}%
\providecommand \BibitemOpen [0]{}%
\providecommand \bibitemStop [0]{}%
\providecommand \bibitemNoStop [0]{.\EOS\space}%
\providecommand \EOS [0]{\spacefactor3000\relax}%
\providecommand \BibitemShut  [1]{\csname bibitem#1\endcsname}%
\let\auto@bib@innerbib\@empty
%</preamble>
\bibitem [{\citenamefont {Shannon}(1948)}]{Shannon}%
  \BibitemOpen
  \bibfield  {author} {\bibinfo {author} {\bibfnamefont {C.~E.}\ \bibnamefont
  {Shannon}},\ }\href@noop {} {\bibfield  {journal} {\bibinfo  {journal} {Bell
  system technical journal}\ }\textbf {\bibinfo {volume} {27}} (\bibinfo {year}
  {1948})}\BibitemShut {NoStop}%
\bibitem [{\citenamefont {Brassard}(2001)}]{brassard2001quantum}%
  \BibitemOpen
  \bibfield  {author} {\bibinfo {author} {\bibfnamefont {G.}~\bibnamefont
  {Brassard}},\ }\href@noop {} {\bibfield  {journal} {\bibinfo  {journal}
  {arXiv preprint quant-ph/0101005}\ } (\bibinfo {year} {2001})}\BibitemShut
  {NoStop}%
\bibitem [{\citenamefont {Buhrman}\ \emph {et~al.}(2010)\citenamefont
  {Buhrman}, \citenamefont {Cleve}, \citenamefont {Massar},\ and\ \citenamefont
  {de~Wolf}}]{buhrman2010nonlocality}%
  \BibitemOpen
  \bibfield  {author} {\bibinfo {author} {\bibfnamefont {H.}~\bibnamefont
  {Buhrman}}, \bibinfo {author} {\bibfnamefont {R.}~\bibnamefont {Cleve}},
  \bibinfo {author} {\bibfnamefont {S.}~\bibnamefont {Massar}}, \ and\ \bibinfo
  {author} {\bibfnamefont {R.}~\bibnamefont {de~Wolf}},\ }\href@noop {}
  {\bibfield  {journal} {\bibinfo  {journal} {Reviews of modern physics}\
  }\textbf {\bibinfo {volume} {82}},\ \bibinfo {pages} {665} (\bibinfo {year}
  {2010})}\BibitemShut {NoStop}%
\bibitem [{\citenamefont {Bennett}\ and\ \citenamefont
  {Brassard}(2014)}]{bennett2014quantum}%
  \BibitemOpen
  \bibfield  {author} {\bibinfo {author} {\bibfnamefont {C.~H.}\ \bibnamefont
  {Bennett}}\ and\ \bibinfo {author} {\bibfnamefont {G.}~\bibnamefont
  {Brassard}},\ }\href@noop {} {\bibfield  {journal} {\bibinfo  {journal}
  {Theoretical computer science}\ }\textbf {\bibinfo {volume} {560}},\ \bibinfo
  {pages} {7} (\bibinfo {year} {2014})}\BibitemShut {NoStop}%
\bibitem [{\citenamefont {Ekert}(1991)}]{ekert1991quantum}%
  \BibitemOpen
  \bibfield  {author} {\bibinfo {author} {\bibfnamefont {A.~K.}\ \bibnamefont
  {Ekert}},\ }\href@noop {} {\bibfield  {journal} {\bibinfo  {journal}
  {Physical review letters}\ }\textbf {\bibinfo {volume} {67}},\ \bibinfo
  {pages} {661} (\bibinfo {year} {1991})}\BibitemShut {NoStop}%
\bibitem [{\citenamefont {Bennett}\ \emph {et~al.}(1993)\citenamefont
  {Bennett}, \citenamefont {Brassard}, \citenamefont {Cr{\'e}peau},
  \citenamefont {Jozsa}, \citenamefont {Peres},\ and\ \citenamefont
  {Wootters}}]{bennett1993teleporting}%
  \BibitemOpen
  \bibfield  {author} {\bibinfo {author} {\bibfnamefont {C.~H.}\ \bibnamefont
  {Bennett}}, \bibinfo {author} {\bibfnamefont {G.}~\bibnamefont {Brassard}},
  \bibinfo {author} {\bibfnamefont {C.}~\bibnamefont {Cr{\'e}peau}}, \bibinfo
  {author} {\bibfnamefont {R.}~\bibnamefont {Jozsa}}, \bibinfo {author}
  {\bibfnamefont {A.}~\bibnamefont {Peres}}, \ and\ \bibinfo {author}
  {\bibfnamefont {W.~K.}\ \bibnamefont {Wootters}},\ }\href@noop {} {\bibfield
  {journal} {\bibinfo  {journal} {Physical review letters}\ }\textbf {\bibinfo
  {volume} {70}},\ \bibinfo {pages} {1895} (\bibinfo {year}
  {1993})}\BibitemShut {NoStop}%
\bibitem [{\citenamefont {Oreshkov}\ \emph {et~al.}(2012)\citenamefont
  {Oreshkov}, \citenamefont {Costa},\ and\ \citenamefont
  {Brukner}}]{oreshkov2012quantum}%
  \BibitemOpen
  \bibfield  {author} {\bibinfo {author} {\bibfnamefont {O.}~\bibnamefont
  {Oreshkov}}, \bibinfo {author} {\bibfnamefont {F.}~\bibnamefont {Costa}}, \
  and\ \bibinfo {author} {\bibfnamefont {{\v{C}}.}~\bibnamefont {Brukner}},\
  }\href@noop {} {\bibfield  {journal} {\bibinfo  {journal} {Nature
  communications}\ }\textbf {\bibinfo {volume} {3}},\ \bibinfo {pages} {1092}
  (\bibinfo {year} {2012})}\BibitemShut {NoStop}%
\bibitem [{\citenamefont {Feix}\ \emph {et~al.}(2015)\citenamefont {Feix},
  \citenamefont {Ara{\'u}jo},\ and\ \citenamefont {Brukner}}]{feix2015quantum}%
  \BibitemOpen
  \bibfield  {author} {\bibinfo {author} {\bibfnamefont {A.}~\bibnamefont
  {Feix}}, \bibinfo {author} {\bibfnamefont {M.}~\bibnamefont {Ara{\'u}jo}}, \
  and\ \bibinfo {author} {\bibfnamefont {{\v{C}}.}~\bibnamefont {Brukner}},\
  }\href@noop {} {\bibfield  {journal} {\bibinfo  {journal} {Physical Review
  A}\ }\textbf {\bibinfo {volume} {92}},\ \bibinfo {pages} {052326} (\bibinfo
  {year} {2015})}\BibitemShut {NoStop}%
\bibitem [{\citenamefont {Gu{\'e}rin}\ \emph {et~al.}(2016)\citenamefont
  {Gu{\'e}rin}, \citenamefont {Feix}, \citenamefont {Ara{\'u}jo},\ and\
  \citenamefont {Brukner}}]{guerin2016exponential}%
  \BibitemOpen
  \bibfield  {author} {\bibinfo {author} {\bibfnamefont {P.~A.}\ \bibnamefont
  {Gu{\'e}rin}}, \bibinfo {author} {\bibfnamefont {A.}~\bibnamefont {Feix}},
  \bibinfo {author} {\bibfnamefont {M.}~\bibnamefont {Ara{\'u}jo}}, \ and\
  \bibinfo {author} {\bibfnamefont {{\v{C}}.}~\bibnamefont {Brukner}},\
  }\href@noop {} {\bibfield  {journal} {\bibinfo  {journal} {Physical Review
  Letters}\ }\textbf {\bibinfo {volume} {117}},\ \bibinfo {pages} {100502}
  (\bibinfo {year} {2016})}\BibitemShut {NoStop}%
\bibitem [{\citenamefont {Chiribella}\ \emph {et~al.}(2008)\citenamefont
  {Chiribella}, \citenamefont {D'Ariano},\ and\ \citenamefont
  {Perinotti}}]{Chiribella}%
  \BibitemOpen
  \bibfield  {author} {\bibinfo {author} {\bibfnamefont {G.}~\bibnamefont
  {Chiribella}}, \bibinfo {author} {\bibfnamefont {G.~M.}\ \bibnamefont
  {D'Ariano}}, \ and\ \bibinfo {author} {\bibfnamefont {P.}~\bibnamefont
  {Perinotti}},\ }\href {\doibase 10.1103/PhysRevLett.101.060401} {\bibfield
  {journal} {\bibinfo  {journal} {Phys. Rev. Lett.}\ }\textbf {\bibinfo
  {volume} {101}},\ \bibinfo {pages} {060401} (\bibinfo {year}
  {2008})}\BibitemShut {NoStop}%
\bibitem [{\citenamefont {Chiribella}\ \emph {et~al.}(2013)\citenamefont
  {Chiribella}, \citenamefont {D’Ariano}, \citenamefont {Perinotti},\ and\
  \citenamefont {Valiron}}]{chiribella2013quantum}%
  \BibitemOpen
  \bibfield  {author} {\bibinfo {author} {\bibfnamefont {G.}~\bibnamefont
  {Chiribella}}, \bibinfo {author} {\bibfnamefont {G.~M.}\ \bibnamefont
  {D’Ariano}}, \bibinfo {author} {\bibfnamefont {P.}~\bibnamefont
  {Perinotti}}, \ and\ \bibinfo {author} {\bibfnamefont {B.}~\bibnamefont
  {Valiron}},\ }\href@noop {} {\bibfield  {journal} {\bibinfo  {journal}
  {Physical Review A}\ }\textbf {\bibinfo {volume} {88}},\ \bibinfo {pages}
  {022318} (\bibinfo {year} {2013})}\BibitemShut {NoStop}%
\bibitem [{\citenamefont {Allen}\ \emph {et~al.}(2016)\citenamefont {Allen},
  \citenamefont {Barrett}, \citenamefont {Horsman}, \citenamefont {Lee},\ and\
  \citenamefont {Spekkens}}]{allen2016quantum}%
  \BibitemOpen
  \bibfield  {author} {\bibinfo {author} {\bibfnamefont {J.-M.~A.}\
  \bibnamefont {Allen}}, \bibinfo {author} {\bibfnamefont {J.}~\bibnamefont
  {Barrett}}, \bibinfo {author} {\bibfnamefont {D.~C.}\ \bibnamefont
  {Horsman}}, \bibinfo {author} {\bibfnamefont {C.~M.}\ \bibnamefont {Lee}}, \
  and\ \bibinfo {author} {\bibfnamefont {R.~W.}\ \bibnamefont {Spekkens}},\
  }\href@noop {} {\bibfield  {journal} {\bibinfo  {journal} {arXiv preprint
  arXiv:1609.09487}\ } (\bibinfo {year} {2016})}\BibitemShut {NoStop}%
\bibitem [{\citenamefont {Procopio}\ \emph {et~al.}(2015)\citenamefont
  {Procopio}, \citenamefont {Moqanaki}, \citenamefont {Ara{\'u}jo},
  \citenamefont {Costa}, \citenamefont {Calafell}, \citenamefont {Dowd},
  \citenamefont {Hamel}, \citenamefont {Rozema}, \citenamefont {Brukner},\ and\
  \citenamefont {Walther}}]{procopio2015experimental}%
  \BibitemOpen
  \bibfield  {author} {\bibinfo {author} {\bibfnamefont {L.~M.}\ \bibnamefont
  {Procopio}}, \bibinfo {author} {\bibfnamefont {A.}~\bibnamefont {Moqanaki}},
  \bibinfo {author} {\bibfnamefont {M.}~\bibnamefont {Ara{\'u}jo}}, \bibinfo
  {author} {\bibfnamefont {F.}~\bibnamefont {Costa}}, \bibinfo {author}
  {\bibfnamefont {I.~A.}\ \bibnamefont {Calafell}}, \bibinfo {author}
  {\bibfnamefont {E.~G.}\ \bibnamefont {Dowd}}, \bibinfo {author}
  {\bibfnamefont {D.~R.}\ \bibnamefont {Hamel}}, \bibinfo {author}
  {\bibfnamefont {L.~A.}\ \bibnamefont {Rozema}}, \bibinfo {author}
  {\bibfnamefont {{\v{C}}.}~\bibnamefont {Brukner}}, \ and\ \bibinfo {author}
  {\bibfnamefont {P.}~\bibnamefont {Walther}},\ }\href@noop {} {\bibfield
  {journal} {\bibinfo  {journal} {Nature communications}\ }\textbf {\bibinfo
  {volume} {6}} (\bibinfo {year} {2015})}\BibitemShut {NoStop}%
\bibitem [{\citenamefont {Rubino}\ \emph {et~al.}(2017)\citenamefont {Rubino},
  \citenamefont {Rozema}, \citenamefont {Feix}, \citenamefont {Ara{\'u}jo},
  \citenamefont {Zeuner}, \citenamefont {Procopio}, \citenamefont {Brukner},\
  and\ \citenamefont {Walther}}]{rubino2017experimental}%
  \BibitemOpen
  \bibfield  {author} {\bibinfo {author} {\bibfnamefont {G.}~\bibnamefont
  {Rubino}}, \bibinfo {author} {\bibfnamefont {L.~A.}\ \bibnamefont {Rozema}},
  \bibinfo {author} {\bibfnamefont {A.}~\bibnamefont {Feix}}, \bibinfo {author}
  {\bibfnamefont {M.}~\bibnamefont {Ara{\'u}jo}}, \bibinfo {author}
  {\bibfnamefont {J.~M.}\ \bibnamefont {Zeuner}}, \bibinfo {author}
  {\bibfnamefont {L.~M.}\ \bibnamefont {Procopio}}, \bibinfo {author}
  {\bibfnamefont {{\v{C}}.}~\bibnamefont {Brukner}}, \ and\ \bibinfo {author}
  {\bibfnamefont {P.}~\bibnamefont {Walther}},\ }\href@noop {} {\bibfield
  {journal} {\bibinfo  {journal} {Science Advances}\ }\textbf {\bibinfo
  {volume} {3}},\ \bibinfo {pages} {e1602589} (\bibinfo {year}
  {2017})}\BibitemShut {NoStop}%
\bibitem [{\citenamefont {Brunner}\ \emph {et~al.}(2014)\citenamefont
  {Brunner}, \citenamefont {Cavalcanti}, \citenamefont {Pironio}, \citenamefont
  {Scarani},\ and\ \citenamefont {Wehner}}]{brunner2014bell}%
  \BibitemOpen
  \bibfield  {author} {\bibinfo {author} {\bibfnamefont {N.}~\bibnamefont
  {Brunner}}, \bibinfo {author} {\bibfnamefont {D.}~\bibnamefont {Cavalcanti}},
  \bibinfo {author} {\bibfnamefont {S.}~\bibnamefont {Pironio}}, \bibinfo
  {author} {\bibfnamefont {V.}~\bibnamefont {Scarani}}, \ and\ \bibinfo
  {author} {\bibfnamefont {S.}~\bibnamefont {Wehner}},\ }\href@noop {}
  {\bibfield  {journal} {\bibinfo  {journal} {Reviews of Modern Physics}\
  }\textbf {\bibinfo {volume} {86}},\ \bibinfo {pages} {419} (\bibinfo {year}
  {2014})}\BibitemShut {NoStop}%
\bibitem [{\citenamefont {Branciard}\ \emph {et~al.}(2015)\citenamefont
  {Branciard}, \citenamefont {Ara{\'u}jo}, \citenamefont {Feix}, \citenamefont
  {Costa},\ and\ \citenamefont {Brukner}}]{branciard2015simplest}%
  \BibitemOpen
  \bibfield  {author} {\bibinfo {author} {\bibfnamefont {C.}~\bibnamefont
  {Branciard}}, \bibinfo {author} {\bibfnamefont {M.}~\bibnamefont
  {Ara{\'u}jo}}, \bibinfo {author} {\bibfnamefont {A.}~\bibnamefont {Feix}},
  \bibinfo {author} {\bibfnamefont {F.}~\bibnamefont {Costa}}, \ and\ \bibinfo
  {author} {\bibfnamefont {{\v{C}}.}~\bibnamefont {Brukner}},\ }\href@noop {}
  {\bibfield  {journal} {\bibinfo  {journal} {New Journal of Physics}\ }\textbf
  {\bibinfo {volume} {18}},\ \bibinfo {pages} {013008} (\bibinfo {year}
  {2015})}\BibitemShut {NoStop}%
\bibitem [{\citenamefont {Almeida}\ \emph {et~al.}(2010)\citenamefont
  {Almeida}, \citenamefont {Bancal}, \citenamefont {Brunner}, \citenamefont
  {Ac{\'\i}n}, \citenamefont {Gisin},\ and\ \citenamefont
  {Pironio}}]{almeida2010guess}%
  \BibitemOpen
  \bibfield  {author} {\bibinfo {author} {\bibfnamefont {M.~L.}\ \bibnamefont
  {Almeida}}, \bibinfo {author} {\bibfnamefont {J.-D.}\ \bibnamefont {Bancal}},
  \bibinfo {author} {\bibfnamefont {N.}~\bibnamefont {Brunner}}, \bibinfo
  {author} {\bibfnamefont {A.}~\bibnamefont {Ac{\'\i}n}}, \bibinfo {author}
  {\bibfnamefont {N.}~\bibnamefont {Gisin}}, \ and\ \bibinfo {author}
  {\bibfnamefont {S.}~\bibnamefont {Pironio}},\ }\href@noop {} {\bibfield
  {journal} {\bibinfo  {journal} {Physical review letters}\ }\textbf {\bibinfo
  {volume} {104}},\ \bibinfo {pages} {230404} (\bibinfo {year}
  {2010})}\BibitemShut {NoStop}%
\bibitem [{\citenamefont {Brunner}\ and\ \citenamefont
  {V\'ertesi}(2012)}]{Vertesi12}%
  \BibitemOpen
  \bibfield  {author} {\bibinfo {author} {\bibfnamefont {N.}~\bibnamefont
  {Brunner}}\ and\ \bibinfo {author} {\bibfnamefont {T.}~\bibnamefont
  {V\'ertesi}},\ }\href {\doibase 10.1103/PhysRevA.86.042113} {\bibfield
  {journal} {\bibinfo  {journal} {Phys. Rev. A}\ }\textbf {\bibinfo {volume}
  {86}},\ \bibinfo {pages} {042113} (\bibinfo {year} {2012})}\BibitemShut
  {NoStop}%
\bibitem [{\citenamefont {Divi\'anszky}\ \emph {et~al.}(2016)\citenamefont
  {Divi\'anszky}, \citenamefont {Trencs\'enyi}, \citenamefont {Bene},\ and\
  \citenamefont {V\'ertesi}}]{Vertesi16}%
  \BibitemOpen
  \bibfield  {author} {\bibinfo {author} {\bibfnamefont {P.}~\bibnamefont
  {Divi\'anszky}}, \bibinfo {author} {\bibfnamefont {R.}~\bibnamefont
  {Trencs\'enyi}}, \bibinfo {author} {\bibfnamefont {E.}~\bibnamefont {Bene}},
  \ and\ \bibinfo {author} {\bibfnamefont {T.}~\bibnamefont {V\'ertesi}},\
  }\href {\doibase 10.1103/PhysRevA.93.042113} {\bibfield  {journal} {\bibinfo
  {journal} {Phys. Rev. A}\ }\textbf {\bibinfo {volume} {93}},\ \bibinfo
  {pages} {042113} (\bibinfo {year} {2016})}\BibitemShut {NoStop}%
\bibitem [{\citenamefont {D\"ur}\ \emph {et~al.}(2000)\citenamefont {D\"ur},
  \citenamefont {Vidal},\ and\ \citenamefont {Cirac}}]{Wstate}%
  \BibitemOpen
  \bibfield  {author} {\bibinfo {author} {\bibfnamefont {W.}~\bibnamefont
  {D\"ur}}, \bibinfo {author} {\bibfnamefont {G.}~\bibnamefont {Vidal}}, \ and\
  \bibinfo {author} {\bibfnamefont {J.~I.}\ \bibnamefont {Cirac}},\ }\href
  {\doibase 10.1103/PhysRevA.62.062314} {\bibfield  {journal} {\bibinfo
  {journal} {Phys. Rev. A}\ }\textbf {\bibinfo {volume} {62}},\ \bibinfo
  {pages} {062314} (\bibinfo {year} {2000})}\BibitemShut {NoStop}%
\bibitem [{\citenamefont {Amico}\ \emph {et~al.}(2008)\citenamefont {Amico},
  \citenamefont {Fazio}, \citenamefont {Osterloh},\ and\ \citenamefont
  {Vedral}}]{Vedral}%
  \BibitemOpen
  \bibfield  {author} {\bibinfo {author} {\bibfnamefont {L.}~\bibnamefont
  {Amico}}, \bibinfo {author} {\bibfnamefont {R.}~\bibnamefont {Fazio}},
  \bibinfo {author} {\bibfnamefont {A.}~\bibnamefont {Osterloh}}, \ and\
  \bibinfo {author} {\bibfnamefont {V.}~\bibnamefont {Vedral}},\ }\href
  {\doibase 10.1103/RevModPhys.80.517} {\bibfield  {journal} {\bibinfo
  {journal} {Rev. Mod. Phys.}\ }\textbf {\bibinfo {volume} {80}},\ \bibinfo
  {pages} {517} (\bibinfo {year} {2008})}\BibitemShut {NoStop}%
\end{thebibliography}%
%\begin{thebibliography}{99}%
%\end{thebibliography}
\end{document}